\documentclass[a4paper]{jpconf}

\usepackage{caption}
\usepackage{subcaption}
\usepackage{graphicx}
\usepackage{url}
\usepackage[colorlinks=false]{hyperref}

\begin{document}

\title{Design and Execution of \texttt{make}-like, distributed Analyses based on Spotify's Pipelining Package Luigi}

\author{M Erdmann, B Fischer, R Fischer, M Rieger}

\address{III. Physics Institute A, RWTH Aachen University, Germany}

\ead{rieger@physik.rwth-aachen.de}

\begin{abstract}
In high-energy particle physics, workflow management systems are primarily used as tailored solutions in dedicated areas such as Monte Carlo production.
However, physicists performing data analyses are usually required to steer their individual workflows manually which is time-consuming and often leads to undocumented relations between particular workloads.
We present a generic analysis design pattern that copes with the sophisticated demands of end-to-end HEP analyses and provides a \texttt{make}-like execution system.
It is based on the open-source pipelining package Luigi which was developed at Spotify and enables the definition of arbitrary workloads, so-called Tasks, and the dependencies between them in a lightweight and scalable structure.
Further features are multi-user support, automated dependency resolution and error handling, central scheduling, and status visualization in the web.
In addition to already built-in features for remote jobs and file systems like Hadoop and HDFS, we added support for WLCG infrastructure such as LSF and CREAM job submission, as well as remote file access through the Grid File Access Library.
Furthermore, we implemented automated resubmission functionality, software sandboxing, and a command line interface with auto-completion for a convenient working environment.
For the implementation of a $t\bar{t}H$ cross section measurement, we created a generic Python interface that provides programmatic access to all external information such as datasets, physics processes, statistical models, and additional files and values.
In summary, the setup enables the execution of the entire analysis in a parallelized and distributed fashion with a single command.
\end{abstract}

\section{Motivation}\label{sec:motivation}

\begin{figure}[h!tbp]
\begin{center}
\includegraphics[width=0.5\textwidth]{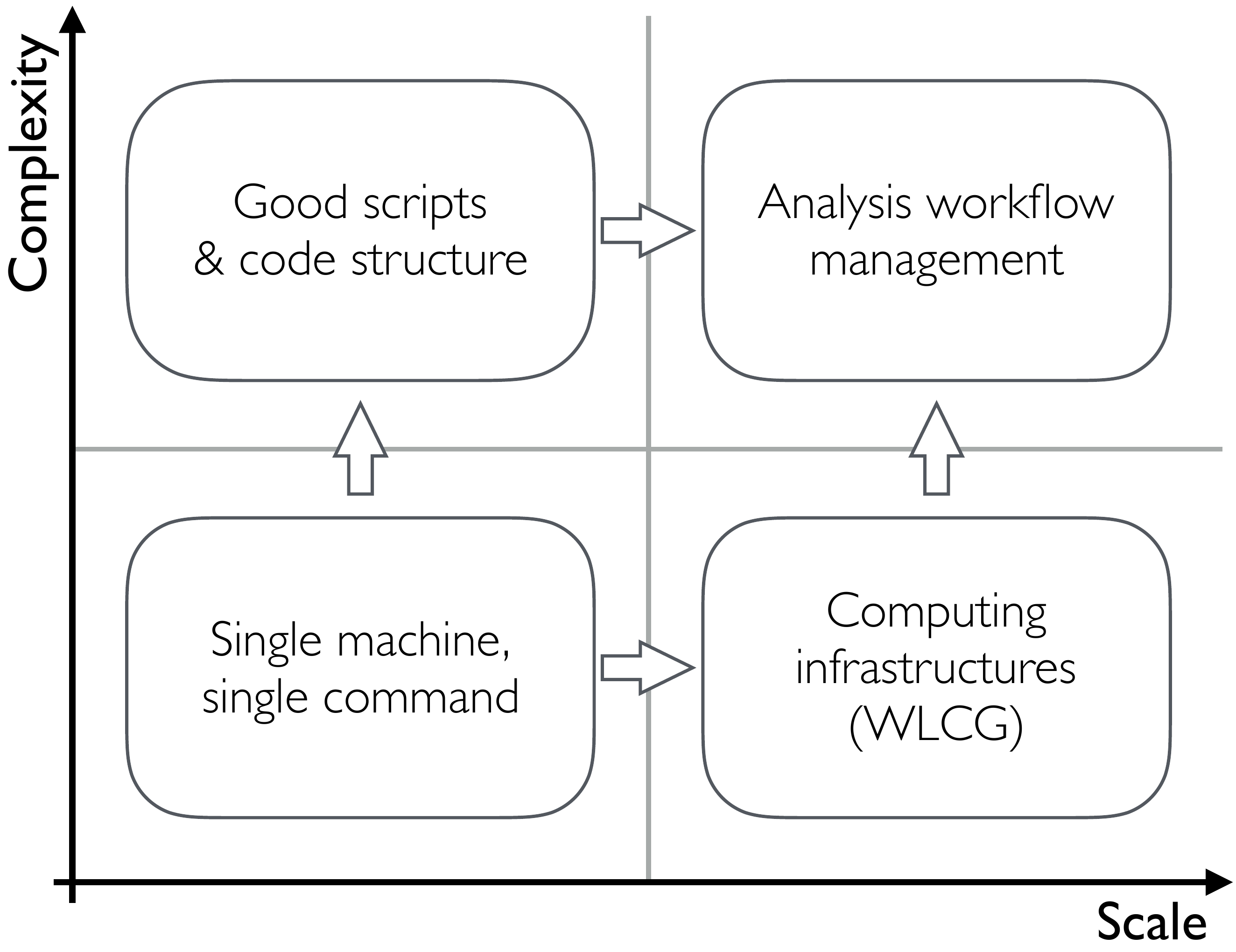}
\caption{Scale and complexity as specification measures for physics analyses and their impact on the choice of structural conception.}
\label{fig:scalecomplexity}
\end{center}
\end{figure}

The management of scientific workflows presents a complex challenge in today's physics working environments in the context of mastering a specific research question.
Current high-energy physics (HEP) analyses are designed to function on a large \textit{scale} as they often require a significant amount of resources in terms of computing time, memory consumption, and disk space.
A logical solution to resource limitation is parallelization on large-scale computing centers.
Main challenges are, e.g., thousands of datasets to be processed, application of multivariate techniques like deep learning, and the diversity of applied algorithms such as final state event reconstruction.
As a consequence, the \textit{complexity} of an analysis, i.e., the degree of granularity and inhomogeneity of workloads, is increased.
The relation between scale and complexity and the impact on analysis conception are indicated in figure \ref{fig:scalecomplexity}.

Physicists are usually required to steer their individual workloads manually, i.e., start (remote) processes, monitor their execution, handle potential failures and resubmission, and eventually initiate dependent workloads once the preceding ones finished successfully.
In addition, intermediate or final output data must be often retrieved for further validation.
On a large scale, this \textit{management} task is not only time-consuming for the operating physicist, but, in contrast to automated approaches, also represents a risk for errors, e.g., the loss of information on the interplay between particular workloads.
In fact, in certain cases the reproducibility of physics results might not be guaranteed.

We present a design pattern for physics analyses conception that copes with the challenges posed by scale and complexity.
Development and testing took place alongside a $t\bar{t}H$ cross section measurement analysis.
Therefore, both its usability and suitability could be demonstrated in a thorough context.
Its core is based on the pipelining package Luigi \cite{luigi}, which provides guidance on structuring arbitrary workloads (section \ref{sec:luigi}).
Scalability on HEP infrastructure is ensured by introducing common interfaces to remote computing facilities (section \ref{sec:submission}) and distributed storage systems (section \ref{sec:storage}).
Furthermore, the portability of software and computing environments via sandboxing is discussed (section \ref{sec:sandboxes}).

\section{Luigi}\label{sec:luigi}

Luigi is a Python module that helps users to ``build complex pipelines of batch jobs, handle dependency resolution, and create visualizations to help manage multiple workflows'' \cite{luigi2}.
The execution model is based on targets and follows a \texttt{make}-like approach as it only computes what is really necessary in order to produce the output of a requested workload \cite{make}.
While its initial development started at Spotify, it was made open-source in 2012, and is now a community-driven project with numerous contributors.

\subsection{Building Blocks}

\begin{figure}[h!tbp]
\begin{subfigure}[b]{0.54\textwidth}
\includegraphics[width=\textwidth]{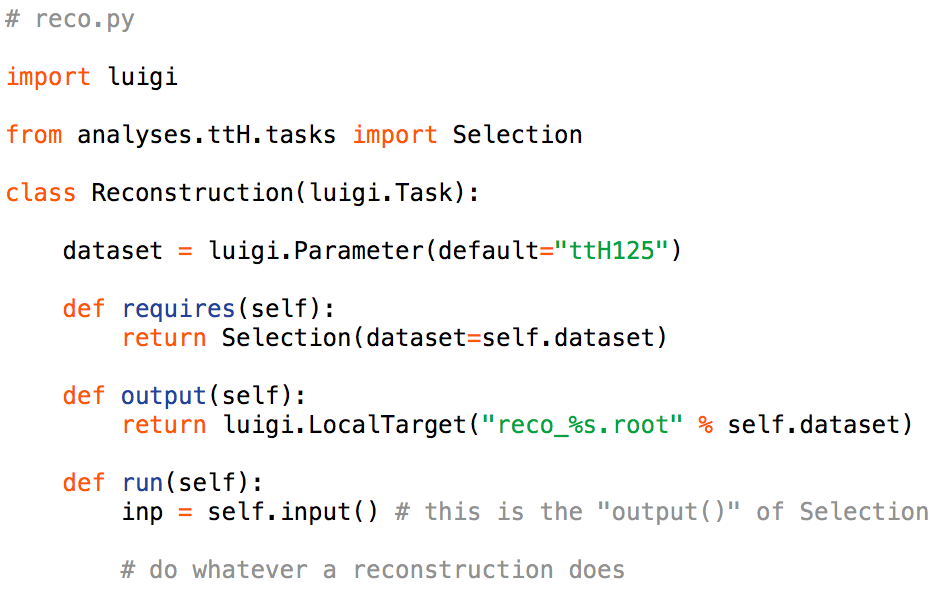}
\subcaption{}
\end{subfigure}
\hfill
\begin{subfigure}[b]{0.45\textwidth}
\includegraphics[width=\textwidth]{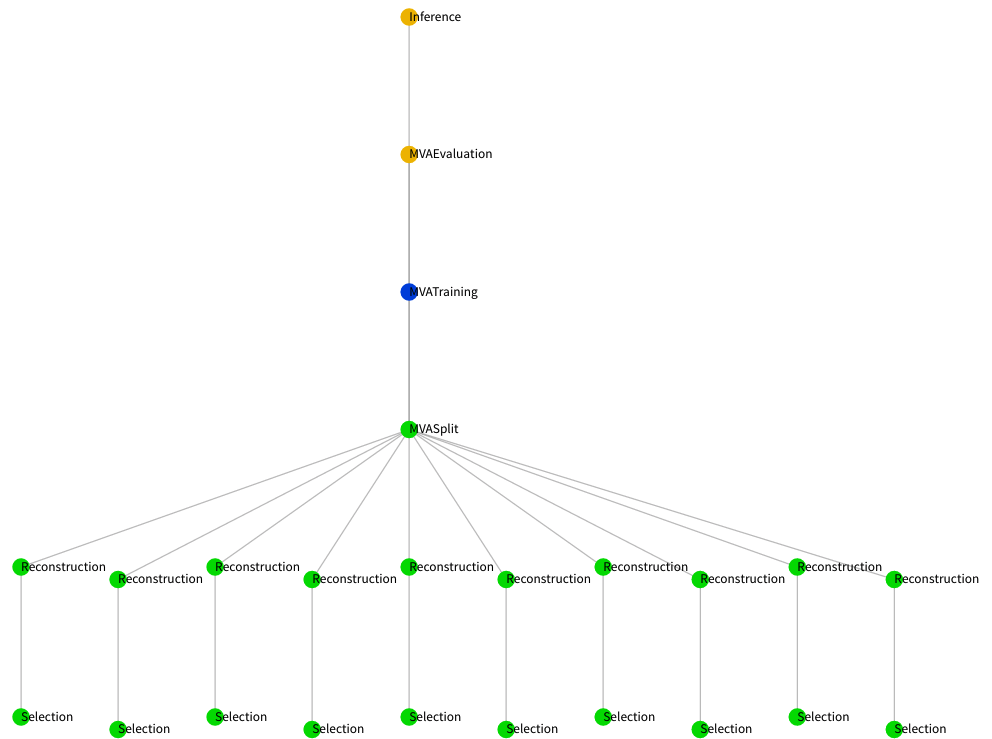}
\subcaption{}
\end{subfigure}
\caption{a) Code example showing the definition of a task class (\texttt{Reconstruction}) with parameters (\texttt{dataset}), required dependencies (\texttt{Selection}), produced output targets (\texttt{LocalTarget}), and a generic payload (\texttt{run}). b) Exemplary task dependency tree. The uppermost task is triggered, all tasks below constitute dependencies.}
\label{fig:task}
\end{figure}

\begin{figure}[h!tbp]
\begin{center}
\includegraphics[width=0.9\textwidth]{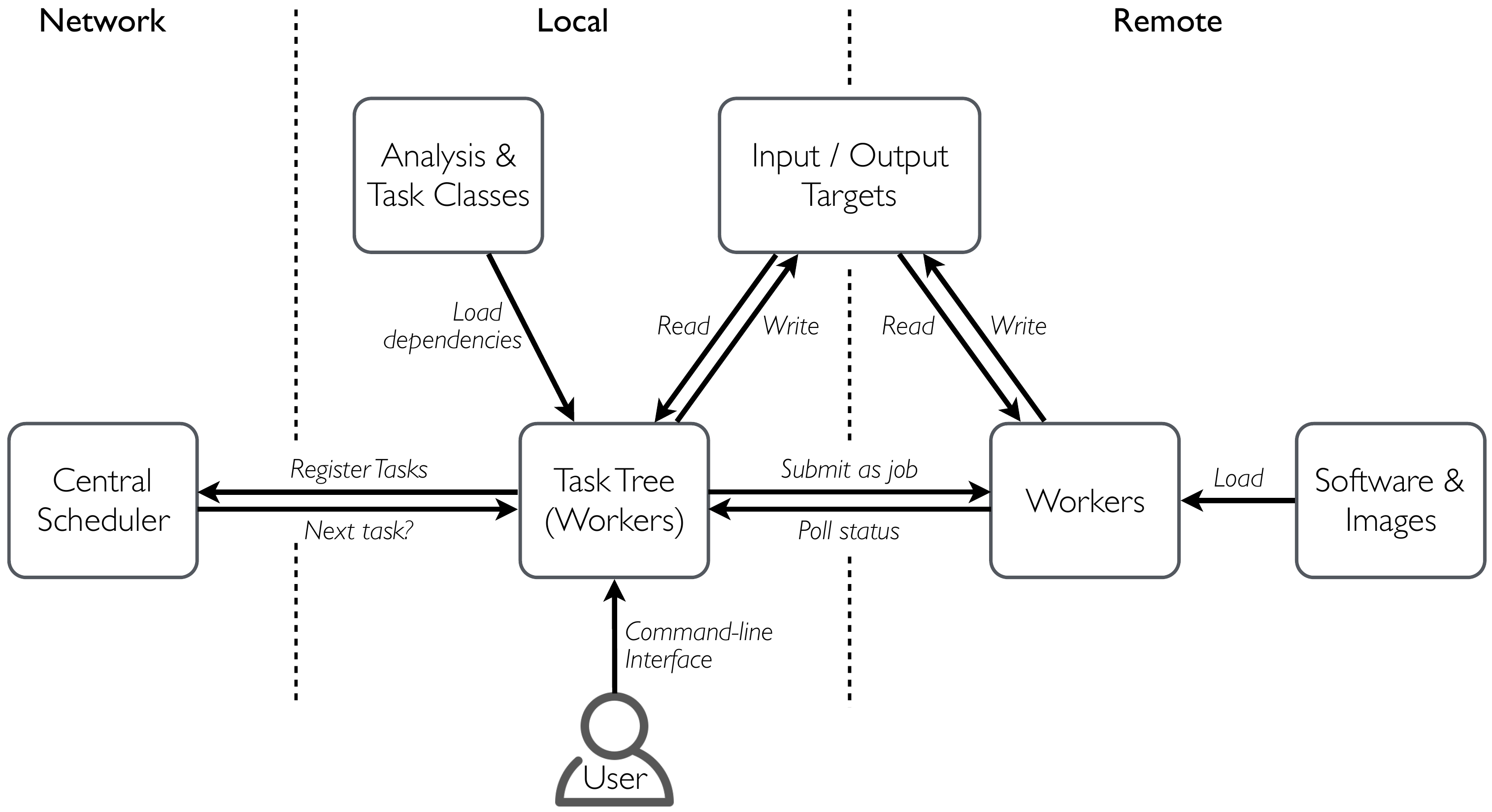}
\caption{Architectural overview showing the interaction between components on local and remote resources. The use of a central scheduler is optional.}
\label{fig:arch}
\end{center}
\end{figure}

Conceptually, Luigi's core functionality is divided into five distinct components represented by Python classes: \texttt{Task}, \texttt{Target}, \texttt{Parameter}, \texttt{Worker}, and \texttt{Scheduler}.
Their implementation is both lightweight and extensible, enabling users to model arbitrary workflows.
Tasks are representations of atomic workload units that constitute a workflow.
They can require one or more other tasks to denote a directional dependency.
The common interface between dependent tasks is accomplished via targets, i.e., containers for arbitrary data such as file paths, database entries, or meta information.
Tasks define their output as a collection of targets that should be created at run time as part of their actual payload.
Parameters can alter the default behavior of tasks, effectively resulting in task \textit{classes} to be considered as \textit{templates}.
They are registered at task definition while their actual value is assigned per task \textit{instance} and can be changed at execution time by the user.
It is often useful to pass parameter values to required tasks and to encode them in output target information, e.g. as path fragment of a file target.
A task is executed in the context of a worker, which holds stateful information such as execution status, run time, and error data.
Workers place tasks in dedicated subprocesses which results in an inherent way of local parallelization.
In addition, workers can communicate with a central scheduler, which not only visualizes worker information and task dependencies in a modern web interface, but also provides a global task lock.
This mechanism prevents situations where, e.g., two users try to run the same task and produce identical output in the same location, which could potentially lead to file corruption.
A code example showing the definition of a simple task class including relevant building blocks and an exemplary dependency tree are presented in figure \ref{fig:task}. An architectural overview scheme of the components is shown in figure \ref{fig:arch}.

Further features include automatic failure handling, command-line interface generation per task, task templates that support map-reduce jobs (e.g. for Hadoop, Pig, or Cascading), and file system abstractions (e.g. for HDFS).

\subsection{Execution Model}

Luigi's execution model follows a \texttt{make}-like approach.
It is initiated by invoking a task with appropriate parameters via the command-line interface or programmatically within Python.
A task is considered complete when all of its output targets exist.
Therefore, the most abstract target definition only contains a single \texttt{exists()} method which evaluates to \texttt{True} if the target exists, or \texttt{False} otherwise.
This rule is initially used to built up a dependency tree.
Starting from the triggered task, all branches of the tree are traversed recursively until a completed task is reached.
Consequently, all tasks collected up to this point will be run using a configurable number of workers, spawned from the process that started the execution.
As a result, Luigi computes only what is really necessary to produce the output of the triggered task.
This paradigm can be classified as output-driven, in contrast to other, workflow-centered approaches where users start workflows as a whole.
Benefits entail automatic output bookkeeping via target existence, transparent and deterministic reproducibility of results, and hence, decoupling of data and algorithms in collaborative working environments.

\section{Remote Computing Resources}\label{sec:submission}

Current high-energy physics data analyses are often comprised of hundreds of thousands computing jobs.
This usually exceeds local resources and makes interfaces to large-scale computing facilities inevitable.
However, typical working environments require users to change existing and/or write additional code to incorporate such interfaces for remote job submission and processing.

We created a mechanism for employing remote computing resources on top of Luigi's task model that follows two paradigms:
\begin{description}
\item 1. Remote computing capabilities of a task should only require minimal, rather descriptive code changes.
\item 2. The decision on the actual run location is not hard-coded, but can be made at execution time.
\end{description}

Technically, this is achieved via \textit{mixin} inheritance of an additional task class that provides submission and status retrieval capabilities.
The rest of the code remains unchanged, although one can optionally define additional requirements that must be met before submission is commenced.
When multiple mixin classes with submission capabilities are given, the particular behavior is steered by a task parameter.
In addition, a user can specify the rules how one or multiple tasks translate into one or multiple jobs.
The remote status is retrieved by a local representation of the corresponding task via polling which can be resumed in case of disruptions.
As an example, we created an implementation for using resources of the Worldwide LHC Computing Grid (WLCG) which constitutes an appropriate use-case in the context of high-energy physics infrastructure.
The mechanism also implements several mandatory features that add actual value to the daily working experience.
They include automatic resubmission for failed jobs, the possibility to define pilot jobs, early stopping criteria with status prediction, and hooks for publishing status information to common job dashboards.

\section{Distributed Data Storage}\label{sec:storage}

The typical disk space consumption of a large-scale physics analyses may amount to multiple tens of gigabytes.
Also, when dealing with remote computing resources and a high degree of parallelization, data should be read from and written to distributed, high-throughput storage systems.

We created a Luigi file system implementation and corresponding target classes that cope with most of the storage back-ends that are deployed in the current high-energy physics landscape.
It is based on Python bindings of the Grid File Access Library (GFAL) \cite{gfal}, which uses plugins to provide support for various back-ends, such as dCache, SRM, GridFTP, XRootD, Amazon S3, WebDAV, and Dropbox \cite{srm,dcache,gridftp,xrootd}.
The implemented interface resembles the local file target implementation, extended by convenience methods for handling file transfers.
User authentication and session management is controlled via environment variables or can be configured programmatically.
Mandatory features include batch transfers, automatic retries for robustness against network and connection disruptions, transfer validation, and local caching to reduce overhead due to redundant requests.

\section{Environment Sandboxing}\label{sec:sandboxes}

Besides portability of specific software environments across remote computing resources, the reproducibility of results over time poses a crucial challenge to the design of physics analyses.
Dependencies on software and data of the host environment should be avoided as they typically are subject to change as part of maintenance and security measures.
Furthermore, a particular workload might require software in an environment that is not compatible with other workloads.
Possible solutions are virtual machines or virtual environments, e.g. via Linux containers, that are to be retained as \textit{images} alongside analysis code \cite{boettiger}.
Virtualization tools like Docker and Vagrant are promising candidates for managing images on a long-term basis \cite{docker,docker3,docker2,vagrant}.

We invented a generalized approach that combines the exchangeability of software environments with Luigi's task and execution model.
So-called \textit{sandboxes} are configured on task level and allow for the definition of rules to either force the switch to a specific sandbox or to identify a fallback sandbox based on runtime conditions.
They are created on demand, host the execution of one ore more tasks, and are terminated automatically once all tasks are completed successfully.
By construction, sandboxes cannot be nested but are rather spawned consecutively by the executing worker process.
This way, each task in a dependency tree can define its own, independent sandbox, which, when left unchanged, leads to reproducible and stable results over time.

\section{Conclusions}\label{sec:conclusions}

The presented tools and concepts for generic analyses conception constitute a novel approach for coping with the increasing demands of modern high-energy physics data analysis.
The Luigi pipelining package is a viable solution to address the complexity of structuring and executing workloads in a \texttt{make}-like fashion.
Scalability in the scope of high-energy physics infrastructure is added in a non-intrusive way by interfacing common job submission systems and remote data storage on arbitrary locations.
In addition, a customizable sandboxing mechanism ensures the integrity of software and computing environments, and therefore the reproducibility of physics results.
Possible implementations are based on virtual environments via Linux containers (Docker) and virtual machines (Vagrant).
As the described approach does not introduce constraints on the software or data structures to be used, it is considered a toolbox providing an \textit{analysis design pattern} rather than a \textit{framework}.

A resulting workflow represents a well-defined formulation of the interplay between particular tasks which often exists only in the ``physicist's head''.
Since all knowledge about the analysis structure is preserved, loss of information is avoided, e.g. in situations when a team or allocations of duties are subject to change.
Furthermore, targets define a flexible but clear interface between tasks which helps to enhance the exchange between individual physicists, small teams or larger groups.
In a broader context, the presented project provides the means to extend the concept of collaboration beyond the sharing of code.
Eventually, the resulting increase of transparency and reproducibility paves the way for \textit{analysis preservation}.

\hfill\\
While started as a private project alongside a $t\bar{t}H$ cross section measurement, the development of all concepts described in this article is being published in the \textit{luigi analysis workflow} project (\url{https://github.com/riga/law}).

\section*{Acknowledgments}

This project is supported by the Ministerium f\"ur Wissenschaft und Forschung, Nordrhein-Westfalen, the Bundesministerium f\"ur Bildung und Forschung (BMBF), the Helmholz Alliance Physics at the Terascale, and the Deutsche Forschungsgemeinschaft (DFG).


\section*{References}

\begin{thebibliography}{9}
\bibitem{luigi} The Luigi Authors, 2011 - 2017, \url{https://github.com/spotify/luigi}

\bibitem{luigi2} Spotify AB, 2012, \textit{Luigi is now open source: build complex pipelines of tasks}, \url{https://developer.spotify.com/news-stories/2012/09/24/hello-world}

\bibitem{make} Free Software Foundation, 2016, \textit{GNU Make}, \url{https://www.gnu.org/software/make}

\bibitem{gfal} The GFAL2 Authors, \textit{Grid File Access Library}, \url{https://dmc.web.cern.ch/projects/gfal-2}

\bibitem{srm} Donno F \textit{et al.}, \textit{Storage resource manager version 2.2: design, implementation, and testing experience}, JPCS, 2008 \textbf{119(6)} 062028

\bibitem{dcache} Millar P \textit{et al.}, \textit{dCache, agile adoption of storage technology}, Computing in High Energy and Nuclear Physics, 2012, New York City (USA), \url{https://bib-pubdb2.desy.de/record/140491}

\bibitem{gridftp} Kourtellis N \textit{et al.}, \textit{Data transfers in the grid: workload analysis of globus GridFTP}, Proceedings of the 2008 International Workshop on Data-aware Distributed Computing, 2008, 29-38, doi:10.1145/1383519.1383523

\bibitem{xrootd} Dorigo A \textit{et al.}, \textit{XROOTD - A highly scalable architecture for data access}, WSEAS Transactions on Computers, 2005


\bibitem{boettiger} Boettiger C, \textit{An introduction to Docker for reproducible research}, ACM SIGOPS Operating Systems Review, Special Issue on Repeatability and Sharing of Experimental Artifacts, 2015 \textbf{49(1)}, 71-79, doi:10.1145/2723872.2723882, \href{https://arxiv.org/abs/1410.0846}{arXiv:1410.0846v1}

\bibitem{docker} Docker Inc., 2016 - 2017, \url{http://www.docker.com}

\bibitem{docker3} Merkel D, \textit{Docker: lightweight Linux containers for consistent development and deployment}, Linux J., 2014 \textbf{239(2)}, \url{http://dl.acm.org/
citation.cfm?id=2600241}

\bibitem{docker2} Matthias K and Kane S P, \textit{Docker: Up and Running: Shipping Reliable Containers in Production}, O'Reilly Media, 2015, ISBN:978-1491917572

\bibitem{vagrant} Hashimoto M, \textit{Vagrant: Up and Running: Create and Manage Virtualized Development Environments}, O'Reilly Media, 2013, ISBN:978-1449335830
\end{thebibliography}
\end{document}